\documentstyle[aps,multicol,epsf,prl]{revtex}

\def\beginwide{
        \end{multicols} \vspace*{-0.5cm} \noindent
        \rule{3.5in}{.1mm}\rule{.1mm}{5mm} \widetext \medskip }
\def\endwide{
        \hspace*{3.5in}~\rule[-5mm]{.1mm}{5mm}\rule{3.5in}{.1mm}
        \begin{multicols}{2}\narrowtext \vspace*{-1.0cm} \noindent }
\def\beginwidetop{
        \end{multicols} \vspace*{-0.5cm} \noindent
        \widetext \medskip }
\def\endwidebottom{
        \begin{multicols}{2} \vspace*{-1.0cm} \noindent }

\begin{document}
\draft
\preprint{draft}

\title{Thermodynamic Fingerprints of Disorder in Flux Line Lattices\\
and other Glassy Mesoscopic Systems}

\author{Thorsten Emig and Mehran Kardar}

\address{Physics Department, Massachusetts Institute of Technology, 
Cambridge, MA 02139, USA}

\date{\today}

\maketitle

\begin{abstract}
We examine probability distributions  for thermodynamic quantities in
finite-sized random systems close to  criticality.
Guided by available exact results, a general ansatz is proposed for 
replicated free energies, which leads to scaling forms for cumulants of 
various macroscopic observables.
For the specific example of a planar flux line lattice in a two dimensional 
superconducting film near $H_{\rm c1}$,
we provide detailed results for the statistics of the magnetic flux density,
susceptibility, heat capacity, and their cross-correlations.
\end{abstract}


\begin{multicols}{2}\narrowtext

Impurities in a sample are expected to modify various measurements,
making it desirable to characterize the probability distribution
functions (PDFs) for the outcomes.  These PDFs may provide important
insight about the underlying physics; as in the case of universal
conductance fluctuations in mesoscopic circuits, a subject of much
recent investigation \cite{Akkermans94}.  Here we consider the
signature of impurities in thermodynamic systems at equilibrium.  At
one extreme, microscopic quantities such as two-point correlation
functions are quite sensitive to disorder; in some cases their PDFs
exhibiting complicated multiscaling behavior\cite{Davis99}.  On the
other hand, the free energy, and other macroscopic properties, are
expected to be self-averaging, converging to fixed values in an
infinite system.  If all correlation lengths in the system are finite,
the PDFs for a mesoscopic (finite-sized) sample should be governed by
the central limit theorem.  We thus focus on systems with long-range
correlations, such as close to a critical point, or in a flux line
(FL) lattice.

In the most interesting cases, {\it disorder is relevant}, leading to
novel correlations distinct from the pure case.  However, due the
difficulty of characterizing collective behavior in such glassy
disordered systems\cite{Nattermann89}, little is known about the
corresponding PDFs.  Aharony and Harris \cite{Aharony96} recently
studied the PDFs of thermodynamic observables, near critical points
with relevant randomness, in finite-sized random systems.  They find a
lack of self-averaging, and universal, non-Gaussian PDFs.  There are
also some results for elastic manifolds pinned by impurities: Using
replicas, Mezard and Parisi relate the scaling of the PDF for
susceptibility with size of a manifold, to its roughness exponent
\cite{Mezard91}.  Exact results on PDFs have been obtained for a
directed polymer on a disordered Cayley tree by Derrida and Spohn,
using a mapping to a deterministic differential equation
\cite{Derrida88}.

An important example of pinned elastic media is provided by FLs in a
superconductor with point impurities.  Indeed, our study was motivated
by the experiment of Bolle {\em et al.} \cite{Bolle99} on a 2d FL
lattice oriented parallel to a thin micrometer-sized film of
2H-NbSe$_2$.  Magnetic response measurements show interesting
sample-dependent fine structure in $B(H)$; a fingerprint of the
underlying pinning landscape.  Interestingly, the problem of 2d lines
with impurities is amenable to an exact solution using replicas, which
gives not only the quenched average but also cumulants of the free
energy\cite{Kardar87}.  We shall use this exact solution to motivate
an ansatz for the scaling of PDFs in general disorder-dominated
thermodynamic systems.

In what follows, we first present a general ansatz for the scaling of
the replicated free energy.  The key assumption is treating the number
of replicas as a scaling field.  Consequences of this ansatz for the
scaling of cumulants of thermodynamic observables in mesoscopic
samples are then enumerated.  We then re-examine the example of FLs in
a 2d layer in more detail, proposing specific experimental tests of
the theoretical picture.

The lack of translational symmetry in the presence of impurities is
cured by calculating disorder averaged moments, $[Z^n]$, of the
partition function.  These moments then provide information about PDFs
of the free energy, magnetization, or susceptibility.  We thus focus
on the scaling of $F_n=-T \ln [Z^n]$, the free energy of $n$ replicas
of the system, with interactions resulting from the disorder average.
In fact $F_n$ can be determined exactly for a single elastic line, or
a lattice of non-crossing lines, in 2d with point impurities
\cite{Kardar87}.  While this is sufficient for the FL experiment
\cite{Bolle99}, we would like to propose a more general scaling ansatz
for random systems.  Consider $np$ replicas of a system of size
$L_\|^{d-1}L_\perp$, with $p$ groups of $n$ replicas subject to the
same reduced temperature $\tau_j$ and scaling field $\psi_j$, for
$j=1\ldots p$.  The vector of scaling fields $\bbox{\psi}$ can consist
of say external magnetic fields, or chemical potentials.  Our scaling
ansatz for the singular part of the free energy density,
$f_p(n,\bbox{\tau},\bbox{\psi})=
-TL_\|^{-d+1}L_\perp^{-1}\ln[\prod_{j=1}^p Z^n(\tau_j,\psi_j)]$, is
\begin{equation}
\label{scal_ansatz}
f_p^{(\rm s)}(n,\bbox{\tau},\bbox{\psi})=b^{-d+1-\zeta}
f_p^{(\rm s)}\left(n b^{\theta_\|}, \bbox{\tau}
b^{1/\nu_\|}, \bbox{\psi} b^{\delta/\nu_\|}\right).
\end{equation}

To include FL systems, we have allowed for anisotropic scaling, with a
`roughness exponent' $\zeta$ relating the $(d-1)$-dimensional
longitudinal ($l_\|$), and the 1d transversal ($l_\perp$) scales, by
$l_\perp \sim l_\|^\zeta$.  (For isotropic systems, we can set
$\zeta=1$.)  Compared to the standard scaling hypothesis for pure
systems, the replica structure introduces $n$ as an additional scaling
field with dimension $\theta_\|=\zeta \theta_\perp$.  The exponent
$\theta_\|$ appears in the modified hyperscaling relation
$2-\alpha=\nu_\|(d-1+\zeta-\theta_\|)$, which follows readily from
Eq. (\ref{scal_ansatz}) by comparing the ${\cal O}(n)$ terms.  Besides
its agreement with available exact results, a mean-field analysis of
the $\phi^4$-model with Gaussian random fluctuations in $T_c$ also
supports our ansatz.

Scaling of disorder averaged cumulants of the thermodynamic
observables can now be extracted from the above ansatz for 
$\ln[\prod_{j=1}^p Z^n(\tau_j,\psi_j)]$. 
Here we enumerate the main
results (details to appear elsewhere):

{\bf (1)} {\em Free energy cumulants}: For $p=1$, we have
\begin{equation}
\ln[Z^n(\tau,\psi)]=\sum_{j=1}^\infty \frac{(-n)^j}{j!}
\frac{[F^j(\tau,\psi)]_c}{T^j},
\end{equation}
where $[\ldots]_c$ denotes the connected or cumulant disorder average.
Using a scaling factor proportional to the correlation length,
$b\sim\xi_\|\sim \tau^{-\nu_\|}$, gives $[F^p(\tau)]_c \sim
\tau^{\nu_\|(d-1+\zeta-p\theta_\|)}$.  For mesoscopic systems of size
$L_\perp\sim L_\|^\zeta$, close to criticality ($\xi_\| > L_\|$) we
must set $b\sim L_\|$, resulting in $[F^p]_c \sim L_\|^{p\theta_\|}$.
The exponent $\theta_\|$ thus characterizes all sample-to-sample
fluctuations in the free energy.

{\bf (2)} {\em Thermal averages}, such as the magnetization or the number
of `particles',  are first order derivatives of the free energy, 
which we will denote generally by $\langle X\rangle
\equiv-\partial F/\partial\psi|_{\psi=0}$.  For a specific realization 
of disorder, the partition function can be expanded as
\begin{equation}
Z(\psi)=Z(0)\sum_{j=0}^\infty \frac{(\psi/T)^j}{j!} \langle X^j
\rangle,
\end{equation}
where $\langle \ldots \rangle$ denotes the thermal average.  Using
this representation, one can show that the $p^{\rm th}$ coefficient of
the expansion of $\ln[Z^n(\psi)]$ with respect to $n\psi$ is the
connected average $[\langle X \rangle^p]_c T^{-p} /p!$.  Applying the
scaling hypothesis of Eq.(\ref{scal_ansatz}), we obtain with
$\delta=2-\alpha-\beta$ the critical scaling behavior $[\langle X
\rangle^p]_c \sim \tau^{\beta_p}$, with
$\beta_p=p\beta-\nu_\|(d-1+\zeta)(p-1)$, and $\beta$ defined by
$[\langle X \rangle] \sim \tau^\beta$.

{\bf (3)} {\em Response functions}: Cumulants of the susceptibility,
$\chi=(\langle X^2 \rangle-\langle X \rangle^2)/T$, can be calculated
in a similar way.  We now consider $p$ sets of $n$ replicas, each with
the same scaling field $\psi_j$, $j=1,\ldots,p$.  The coefficient of
the term $n^p(\psi_1\cdot\ldots\cdot\psi_p)^2$ of the expansion of the
generating function $\ln[\prod_{j=1}^p Z^n(\psi_j)]$ is given by the
cumulant average $(2T)^{-p}[\chi^p]_c$.  This gives a critical scaling
of the form $[\chi^p]_c \sim \tau^{-\gamma_p}$, with
$\gamma_p=p\gamma+\nu_\|(d-1+\zeta)(p-1)$.

{\bf (4)} {\em Cross correlations}, such as $\Gamma_{XY}=\langle XY
\rangle - \langle X \rangle \langle Y \rangle$, where $Y$ is a
derivative of the free energy with respect to another scaling field
$\tilde\psi$, are also of interest.  For example, the cross
correlation, $\Gamma_{EM}$, of magnetization $M$ and energy $E$, has
been measured in numerical simulations, since it allows for a more
accurate estimate of the exponent ratio $\alpha/\nu$ then a direct
method \cite{Heuer90}.  The generator of $\Gamma_{XY}$ is
$\ln[\prod_{j=1}^p Z^n(\psi_j,\tilde\psi_j)]$, with coefficients
$T^{2p}[\Gamma_{XY}^p]_c$ corresponding to the terms
$n^p\psi_1\tilde\psi_1\ldots\psi_p\tilde\psi_p$.  Our scaling ansatz
yields $\Gamma_{EM}\sim\tau^{p(\beta-1)-\nu_\|(d-1+\zeta)(p-1)}$.
There are also correlations between different susceptibilities.  Along
the lines presented above, it is possible to show that these
cross-correlations can be generated from the expansion of
$\ln[\prod_{j=1}^p Z^n(\psi_j)Z^n(\tilde\psi_j)]$, and that the
cumulant $(2T)^{-2p}[(\chi\tilde\chi)^p]_c$ is the coefficient of the
term $n^{2p}\psi_1^2\tilde\psi_1^2\ldots\psi_p^2\tilde\psi_p^2$.
Choosing for $\chi$ and $\tilde\chi$ the magnetic susceptibility and
the heat capacity $c$, respectively, we get $[(\chi c)^p]_c \sim
\tau^{-2\gamma_p-\nu_\|(d-1+\zeta)}$.

Finally, we compare with the results of Aharony and
Harris\cite{Aharony96}, on the relative cumulants
$R_{p,X}=[X^p]_c/[X]^p$, in a $\phi^4$-model with random $T_c$.  By a
perturbative renormalization group (RG), in which they assume a
Gaussian distribution for randomness on {\em all} length scales, they
obtain $R_{p,\chi}=p!2^{p-3}R_{2,\chi}^{p-1}$, for the magnetic
susceptibility.  Indeed, for all observables (including
susceptibilities) we find $R_{p,X}\sim R_{2,X}^{p-1}$, but the exact
coefficient cannot be obtained from the scaling ansatz. However, even
for an originally Gaussian random $T_c$, higher then second cumulants
are generated by RG transformations and yield additional (universal)
contributions to the above coefficient \cite{Aharony96}.  Assuming
that randomness is relevant at the phase transition, we find for the
second relative cumulant of any observable $X$ that $R_{2,X}\sim
(b/L_\|)^{d-1} (b^\zeta/L_\perp)$.  Choosing a scaling factor $b\sim
\xi_\|$ if $\xi_\| \ll L_\|$, and $b\sim L_\|$ in the critical regime
$\xi_\| \gg L_\|$, we reproduce the results of Ref. \cite{Aharony96}.

We now apply these results to the experimental study by Bolle {\em et
al.} of a planar, randomly pinned vortex array penetrating a
mesoscopic quasi $2$-dimensional thin single crystal of 2H-NbSe$_2$
with weak pinning \cite{Bolle99}.  The crystal was glued onto a
silicon micromachined mechanical resonator.  By measuring jumps in the
resonant frequency caused by magnetic FLs entering the sample, the
number of lines was determined very accurately as a function of the
applied magnetic field.  Close to the lower critical field $H_{\rm
c1}$, the jumps were irreversible, indicative of the difficulty of FLs
finding optimal pinning configurations.  At higher fields, the
increased line density should result in a smaller pinning length,
enabling the FL lattice to find its optimal pinning state.  Indeed
reversible behavior is observed experimentally in this case.  In both
regimes, the response depends on the detailed configuration of pinning
sites, the sample geometry, and the vortex interactions.  Therefore,
the observed discrete jumps in the constitutive relation, $B(H)$,
provide a fingerprint of the disorder in a specific sample.

Previous theoretical and experimental work concentrated mainly on the
determination of the relation $B(H)$ near $H_{\rm c1}$ for different
kinds of disorder \cite{Lehrer99}.  Fluctuations in $B(H)$ are also of
interest, characterizing more clearly the non-trivial effects of
randomness in this mesoscopic system.  Therefore, we study entire PDFs
for observables like $B$, magnetic susceptibility $\chi$, heat
capacity $c$, and correlations between them.  First consider a single
FL to introduce the basic notations.  The transversal wandering of
such a line is described by a trajectory $x(y)$, which has an energy
\cite{Nelson88}
\begin{equation}
{\cal H}=\int_0^L dy \left[ \frac{g}{2} \left(\frac{dx}{dy} \right)^2 + V(x(y),y)\right].
\end{equation} 
Here, $g=\phi_0 H_{\rm c1}/(4\pi)$ is the elastic stiffness of the FL
carrying a flux quantum $\phi_0$.  The random potential $V(x,y)$
mimics point like pinning, and is characterized by the disorder
average $[V(x,y)V(x',y')]=\Delta\delta_\xi(x-x')\delta(y-y')$, with
strength $\Delta$ and a short range function, $\delta_\xi$, of width
of the in-plane coherence length $\xi$.  The dimensions of the sample
are taken to be $L$ along the field direction $\hat{\bf y}$, and $W$
in the transverse direction $\hat{\bf x}$.

A single FL can wander freely in the transverse direction to take
advantage of the randomly distributed pinning centers.  This leads to
an anomalous growth of its displacement by $\delta x(L) \sim
L^{2/3}$\cite{Huse85}.  In contrast, in a lattice of lines, the
non-crossing condition is a strong restriction for the possible
configurations of each line.  The results of Ref. \cite{Kardar87} for
such a lattice, only apply to FLs at temperatures $T$ larger than
$T^*=(g\xi\Delta)^{1/3}$, the height of the smallest energy barrier
due to the random potential.  Recently, Korshunov and Dotsenko
\cite{Korshunov98} generalized the results for a single line to the
low temperature limit, $T \ll T^*$, by using a replica interaction
potential with a small but finite curvature, instead of the
rectangular well which works in the high-$T$ limit.  The
generalization of their solution to an array of lines is
straightforward.  The non-trivial part of the free energy of $n$
replicas of the system, with a fixed number of $N=W/a_0$ of lines at
mean separation $a_0$, can be summarized in both limits as
\cite{Kardar87}
\begin{equation}
\label{F_n}
\frac{F_n(N)}{T}=-\ln [Z^n]=nkN^2 \frac{L}{W} \frac{\Delta}{T^2}\, 
{\cal G}\left( n\sqrt{\frac{kg\Delta W}{NT^3}}\right),
\end{equation}
with the parameter $k=1$ for $T \gg T^*$, and $k\approx T/T^*$ for
$T\ll T^*$.  Here we have neglected the trivial contribution to the
free energy which is linear in $N$, and ${\cal G}$ is an analytic
function \cite{note1}.

To study the fluctuations in the number of FLs near $H_{\rm c1}$, we
use a grand canonical description with a chemical potential $\mu=ghL$,
where $h=(H-H_{\rm c1})/H_{\rm c1}$ is the reduced magnetic field.
The PDF of $N$ is characterized by the disorder averaged cumulants
$[N^p]_c$ in the number of FLs, which are given by $T^p$ times the
coefficients of the terms $(n\mu)^p$ in the expansion of $\ln
[Z^n(\mu)]$, leading to
\begin{equation}
\label{N_moments}
[N^p]_c\sim WL^{1-p} \left(\frac{gT}{k\Delta}\right)^{2-p} h^{(5-3p)/2}.
\end{equation}
The moments of the magnetic flux density follow from $B=\phi_0
N/(dW)$, where $d$ is the thickness of the sample.  It is interesting
to note that the second cumulant is {\em universal}, $[N^2]_c \sim
(W/L) h^{-1/2}$, independent of temperature and disorder strength, in
both the high and low temperature regimes.  For $T\ll T^*$, all
cumulants are independent of $T$, but do depend on the disorder
strength.  In the thermodynamic limit $WL\to\infty$, all cumulants
$[B^p]_c$ approach zero for $p>1$.  On the other hand, for mesoscopic
systems, the divergence of $[B^p]_c$ for $h\to 0$ is stopped if the
average distance $a_0$ between the FLs approaches the system size $W$.
Deep in the glassy low density phase near $H_{\rm c1}$, we obtain from
Eq. (\ref{F_n}) for the free energy cumulants
\begin{equation}
[F^p]_c\sim WL g^2 \left(\frac{T}{k\Delta} \right)^{2-p} h^{(5-p)/2}.
\end{equation}
This agrees with our scaling ansatz with exponents $\zeta=2/3$,
$\theta_\|=1/3$, $\nu_\|=3/2$ and $\delta=1$, and gives a universal
value for $p=2$, which still depends on the stiffness $g$.

Now consider the response of the FL lattice to changes in the external
magnetic field, measured by the susceptibility $\chi$.  This is
related to thermal fluctuations in the number of FLs by
$\chi=(L/W)(\langle N^2 \rangle -\langle N\rangle^2)/T$, for a fixed
realization of disorder.  The disorder averaged $p^{\rm th}$ cumulant
of the number fluctuations can be obtained from the generating
function $\ln [ \prod_{j=1}^{p} Z^n(\mu_j) ]$ for $p$ different
chemical potentials $\mu_j$, giving the susceptibility cumulants
\begin{equation}
\label{chi_moments}
[\chi^p]_c \sim (WL)^{1-p} g^{2-2p}  \left(\frac{T}{k\Delta} \right)^{2-p} 
h^{-5(p-1)/2}.
\end{equation}
This result deserves a few comments: First, the disorder averaged
susceptibility $[\chi]$ is non-singular at the transition $h\to 0$,
since the exponent vanishes for $p=1$.  In fact, due to a statistical
tilt symmetry\cite{Hwa94} the susceptibility is simply related to the
compression and tilt elastic moduli ($c_{11}$, $c_{44}$) of the FL
lattice, by $[\chi]=(2\pi/a_0)^2(c_{11}c_{44})^{-1/2}$.  This is in
agreement with our result $[\chi]\sim T/(k\Delta)$, as can be seen by
using $c_{44}=g/a_0$, rewriting $c_{11}$ in terms of the steric
repulsion between the lines, and estimating $a_0=W/N$ from
Eq. (\ref{N_moments}) with $p=1$, and a value for $h$ of order one.
Again, the variance is the only moment of the susceptibility, which
shows universality, $[\chi^2]_c \sim (WLg^2)^{-1} h^{-5/2}$
\cite{note2}.

Next, consider the response of a fixed number of FLs to changes in
temperature.  Keeping the number $N$ of FLs constant by adjusting the
magnetic field, but changing the temperature, allows us to study
thermal fluctuations around the global ground-state.  Physically, the
response of the FL lattice to changes in magnetic field $H$, or
temperature $T$, are quite different: A change in $H$ (or $N$) usually
leads to a complete rearrangement of the whole ensemble of lines.
Increasing $T$, however, causes stronger entropic repulsions between
the $N$ lines, which now fluctuate around their state of optimal
pinning.  The response to a change in $T$ also depends on the detailed
pinning landscape, producing a heat capacity $c$ which is sample
specific.  Therefore, the statistics of $c$ are also of interest; its
cumulants can be obtained from the generating function $\ln
[\prod_{j=1}^{p} Z^n(\beta_j) ]$ with $p$ different temperatures
$T_j=1/\beta_j$, as (for fixed $N$)
\begin{equation}
\label{c_moments}
[c^p]_c\sim LW^{\frac{p-3}{2}} (k\Delta)^{\frac{p+1}{2}} g^{\frac{p-1}{2}}
N^{\frac{5-p}{2}} T^{-\frac{3p+1}{2}}.
\end{equation}
In the high temperature limit ($T \gg T^*$), the moments $[c^p]_c\sim
T^{-(3p+1)/2}$ decay faster with temperature than in the low
temperature limit ($T \ll T^*$), where $[c^p]_c\sim T^{-p}$.

We may also examine cross-correlations between the different
susceptibilities such as the heat capacity $c$, and the magnetic
susceptibility $\chi$.  In the thermodynamic limit, i.e.  beyond a
characteristic system size, different susceptibilities are expected to
be statistically independent.  For example, this has been demonstrated
for the magnetic susceptibilities of two noninteracting FL lattices
with different random potentials\cite{Hwa94}.  The correlations
between $c$ and $\chi$ can be obtained from our scaling theory by
extracting the coefficients of $n^{2p} \mu_1^2\beta_1^2
\ldots\mu_p^2\beta_p^2$ of $\ln[\prod _{j=1}^p Z^n(\mu_j) \prod
_{j=1}^p Z^n(\beta_j)]$.  Using the results of
Eqs. (\ref{chi_moments}) and (\ref{c_moments}), we get
\begin{equation}
\frac{[(\chi c)^p]_c}{[\chi^p]_c [c^p]_c} \sim (WL)^{-1} \left( 
\frac{k\Delta}{gT}\right)^2 h^{-5/2}.
\end{equation}
For finite systems, the divergence as $h \to 0$ is cut off at $h\sim
k\Delta/(gTW)$.  Then the result can be rewritten as $[(\chi
c)^p]_c/([\chi^p]_c [c^p]_c) \sim L_c/L$, with a characteristic length
scale $L_c=(gT/k\Delta)^{1/2} W^{3/2}$ for the decay of correlations.
This length-scale has a simple physical interpretation: $L_c$ is the
length of a single FL whose transverse wanderings can explore the
whole sample width $W$.  Since the transversal fluctuations of a line
of length $L$ is grow as $\delta x(L) \approx (k\Delta/gT)^{1/3}
L^{2/3}$\cite{Korshunov98}, we get the above result for $L_c$ from the
condition $\delta x(L_c)\approx W$.

The full shape of the PDFs is determined by the relative cumulants
$R_{p,X}=[X^p]_c/[X]^p$.  For $X=N$, $\chi$, $c$ or $\chi c$, our
results yield $R_{p,X} \sim R_{2,X}^{p-1}$ up to a numerical
coefficient.  Therefore, the system parameters enter the PDF shapes
only through $R_{2,X}$, which is interestingly independent of the
observable $X$, and given by
\begin{equation}
R_{2,X}\sim (WL)^{-1} \left( 
\frac{k\Delta}{gT}\right)^2 h^{-5/2}  = \frac{\xi_\perp\xi_\|}{WL}.
\end{equation}
The length scales in the final expression are $\xi_\perp=a_0$, the
separation of FLs, and $\xi_\|= (gT/k\Delta)^{1/2} a_0^{3/2}$, the
mean longitudinal distance between collisions of FLs.  To obtain a
size independent PDF, anisotropy requires looking at finite-size
samples of width $W\sim L^\zeta$, with roughness exponent $\zeta=2/3$.
The resulting PDF is however, still {\em non-universal}, depending on
$R_{2,X}\sim (gT/k\Delta)^{1/2}$.

The above predictions for the scaling of PDFs, or cumulants, can be
tested experimentally by measurements on different realizations of
randomness, drawn from the same distribution of impurities.
Generating many such different realizations could in fact be quite
easy, depending on the system under study.  For example, in the case
of the FL lattice of Ref.\cite{Bolle99} experiments can be performed
on the same sample, with different realizations of randomness
generated by simply rotating the sample with respect to the external
magnetic field.  Each (finite-size) realization of randomness yields a
characteristic value for thermodynamic observables, providing a
reproducible `fingerprint' of the configuration.  The statistics of
these measurements is then described by the cumulants calculated here.

We thank D.R. Nelson for bringing the experimental system of
Ref.\cite{Bolle99} to our attention.  This work was supported by the
Deutsche Forschungsgemeinschaft under grant EM70/1-1 (TE), and the
National Science Foundation grant No.  DMR-98-05833 (MK).

\end{multicols}


\begin{references}

\bibitem{Akkermans94} {\em Mesoscopic Quantum Physics} edited by
E. Akkermans {\em et al.} (North-Holland, Amsterdam, 1994).

\bibitem{Davis99} T. Davis and J. Cardy, preprint cond-mat/9911083.

\bibitem{Nattermann89} T. Nattermann, Int. J. Mod. Phys. B {\bf 3},
1597 (1989).

\bibitem{Aharony96} A. Aharony, A.B. Harris,  Phys. Rev. Lett. {\bf
77}, 3700 (1996).

\bibitem{Mezard91} M. Mezard and G. Parisi, J. Phys. I (Paris) {\bf 1}, 
809 (1991).

\bibitem{Derrida88} B. Derrida and H. Spohn, J. Stat. Phys. {\bf 51}, 
817 (1988).

\bibitem{Bolle99} C.A. Bolle {et al.}, Nature {\bf 399}, 43 (1999).

\bibitem{Kardar87} M. Kardar, Nucl. Phys. B {\bf
290}, 582 (1987).

\bibitem{Heuer90} H.O. Heuer, Phys. Rev. B {\bf 42}, 6476 (1990).

\bibitem{Lehrer99} For a recent detailed summary of the theoretical
results see: R.A. Lehrer and D.R. Nelson, preprint cond-mat/9908117.

\bibitem{Nelson88} D.R. Nelson, Phys. Rev. Lett. {\bf 60}, 1973
(1988); T. Nattermann and R. Lipowsky, Phys. Rev. Lett. {\bf 61}, 2508
(1988).

\bibitem{Huse85} D.A. Huse, C.L. Henley, and D.S. Fisher, Phys. Rev. Lett. 
{\bf 55}, 2924 (1985).

\bibitem{Korshunov98} S.E. Korshunov, V.S. Dotsenko, J. of
Phys. A {\bf 31} 2591 (1998).

\bibitem{note1} The result of Eq. (\ref{F_n}) is valid for systems of size $W 
\gg T^3/(k^3g\Delta)$. For 2H-NbSe$_2$, in the relevant regime of $T\ll T^*$, this 
condition becomes $W\gg \xi$, which is fulfilled even for mesoscopic samples.

\bibitem{Hwa94} T. Hwa, D.S. Fisher, Phys. Rev. Lett. {\bf 72}, 2466 (1994).

\bibitem{note2} Universal sample-to-sample fluctuations of the magnetic 
susceptibility have been predicted in this 2d FL system at the
so-called glass transition in Ref. \cite{Hwa94}, and were confirmed by
numerical simulations in C. Zeng, P.L. Leath, and T. Hwa,
Phys. Rev. Lett. {\bf 83}, 4860 (1999).

\end{references}
\end{document}